\documentclass[12pt]{article}
\usepackage{geometry}                % See geometry.pdf to learn the layout options. There are lots.
\usepackage{graphicx}
\usepackage{amssymb}

\title{Application of the Bethe-Salpeter Equation to Mesonic Atoms}
\author{D. A. Owen\footnote{Physics Department, Ben Gurion Uinversity, Beer Sheva, Israel} \and Roger C. Barrett\footnote{Physics Department, University of Surrey, Guildford, Surrey, U.K.}}
\begin{document}
\maketitle
\begin{abstract}
We use quantum electrodynamics and the Bethe-Salpeter equation to calculate the bound state energies for a two-particle system comprised of a spin-0 and spin-1/2 particle.  We generalize our treatment to include the finite size of the nucleus and of the meson and our results can be applied to kaonic and pionic atoms. Our treatment includes quantum electrodynamics and recoils corrections to all orders.
\end{abstract}
 
 \section{Introduction}
 Exotic atoms can be of great value in the investigation of the threshold of the onset of strong interactions.  For this it is useful to use the difference between the measured X-ray energy and the value of the corresponding electromagnetic energy\cite{deloff}.  Thus to gain insight into the strong interaction effects, we must remove the electromagnetic contribution so we need to know how to calculate this precisely. In 1990 Owen found a new representation of the Klein-Gordon propagator which allows a convenient formulation of the bound state problem using the Bethe-Salpeter equation, or which one of the particles is spinless \cite{owen} \cite{halpert} \cite{foundations}.  By separating the positive and negative energy parts of the propagator, a two-component formalism was obtained in which the inverse of the Klein-Gordon propagator is linear in the momentum $p_{0}$.  This makes it possible to combine it with the Dirac propagator, and  generalize this formalism to include the finite size (i.e., form factors) of both the particles.  Thus our aim is to calculate the energy levels of the meson-nucleus system with a purely electromagnetic interaction which includes all recoil corrections.  This necessitates the use of the Bethe-Salpeter equation for systems in which recoil cannot be neglected.  In the Bethe-Salpeter equation the entire recoil contribution to the energy is included.  For S-states the effect of the strong interaction is dominant  and a comparisons with experiment cannot be made until this is taken into account.  We leave these considerations for a future work since our concern here is the purely electromagnetic interaction.
 
There has been interest in this problem due to new theoretical treatment using the Breit equation \cite{kelbar} and recent experimental results.  We must point out that there is a fundamental difference between the Breit equation approach, which is based on an approximate formulation of the two body problem, and the Bethe-Salpeter equation which is a method for finding the location of the poles of the two-particle, relativistic quantum Green's function.  As it relates to quantum electrodynamics, this approach can give the location of the energy levels as precisely as one desires.

\section{Review of the Theory}
\subsection{Lagrangian}
 We take for our free Lagrangian \cite{halpert}  \cite{foundations}
 
 \begin{equation}
 \mathcal{L}_{0}=\bar{\psi}'(\not p-m_{a})\psi'+\frac{i}{2}(\bar{\Phi}'\ \dot{\Phi}'-\dot{\bar{\Phi}}'\Phi')-\sum_{k=1}^{3}\frac{\partial \bar{\Phi}'}{\partial x^{k}}\frac{M}{2m_{b}}\frac{\partial \Phi'}{\partial x^{k}}-m_{b}\bar{\Phi}'\beta_{s}\Phi'
  \end{equation}
 while our interaction Lagrangian is obtained from Eq. 1 by minimal coupling and M and $\beta_{s}$ are given by the matrices, respectively:
 \begin{equation} 
 M=\left(\begin{array}{rr} 1& 1 \\ -1 & -1 \end{array}\right) \ \ \ \ \ \ \ \ \ \ \ \ \ \ \ \ \beta_{s}=\left(\begin{array}{rr} 1 & 0 \\0 & -1 \end{array}\right)
 \end{equation}
 while $m_{a}$ and $m_{b}$ are the respective masses of the particles.  The total Lagrangian for our nucleon-meson system can be written
 \begin{equation}
 \mathcal{L}_{0-1/2}=\mathcal{L}_{0}-\frac{1}{4}F^{\mu \nu}F_{\mu \nu}+\mathcal{L}^{int}_{0-1/2}
 \end{equation}
 where the second term is the Lagrangian of the free electro-magnetic field and third term represents the electromagnetic interaction between the nucleon and meson.  The hadronic content will be expressed by the charge distribution of these particles.  That is, we can write \cite{drell}
 
 \begin{eqnarray}  \label{intH1}
 \mathcal{H}^{int}_{0-1/2}(x)=-\mathcal{L}^{int}_{0-1/2}=ie_{b}\mathbf{A}'(\mathbf{x}',t)\rho_{\pi}(\mathbf{x}'-\mathbf{x})\cdot \left[(\nabla\bar{\Phi}'(x))\frac{M}{2m_{b}}\Phi'(x)-\bar{\Phi}'(x))\frac{M}{2m_{b}}\nabla\Phi'(x)\right] \nonumber \\ +e_{b}^{2}\mathbf{A}'^{2}(\mathbf{x}',t)\rho_{\pi}(\mathbf{x}'-\mathbf{x})\bar{\Phi}'(x)\frac{M}{2m_{b}}\Phi'(x)-e_{a}\psi'^{+}(x)\gamma \cdot \mathbf{A}'(\mathbf{x}',t)\rho_{N}(\mathbf{x}'-\mathbf{x})\psi'(x)+\mathcal{H}_{c}(x)
   \end{eqnarray}
 where $\rho_{\pi}$ and $\rho_{N}$ are the Fourier transforms of the pion and nucleon and where the contribution to the Coulomb Hamiltonian is given by
 \begin{equation} \label{coulomb}
 H_{c}=\int d^{3}x \mathcal{H}_{c}(x)=H_{cs}+\frac{e_{a}e_{b}}{4\pi}\int d^{3}xd^{3}y \frac{\psi'^{+}(\mathbf{x},t)\psi'(\mathbf{x},t)\bar{\Phi}'(\mathbf{y},t)\Phi'(\mathbf{y},t)\rho_{N}(\mathbf{x}-\mathbf{x}')\rho_{\pi}(\mathbf{y}-\mathbf{y}')}{|\mathbf{x}'-\mathbf{y}'|}
 \end{equation}
 where the integrations over $x'$ and $y'$ are understood and $H_{cs}$ describes the Coulomb self-interaction Hamiltonian which only plays apart when renormalization is considered.  In principle, we should have added an additional term to Eq. \ref{intH1} to describe the contribution of anomalous magnetic moment of the proton to the interaction Hamiltonian. However, in the bound state energy levels that we consider, this term will not give any additional contribution so for simplicity, we have neglected it. 
 
\subsection{Bethe-Salpeter Equation}
The Bethe-Salpeter equation \cite{Bethe} can be written as
\begin{equation}  \label{BSeq}
G(1,2;3,4)=G_{0}(1,2;3,4)+G_{0}(1,2;5,6)I(5.6;7,8)G(7,8;3,4)
\end{equation}
 where
\begin{equation}
G(1,2;3,4)=-<0|T[\psi'(x_{1})\Phi'(x_{2})\bar{\psi}'(x_{3})\bar{\Phi}'(x_{4})]|0>=-<0|T[\psi(x_{1})\Phi(x_{2})\bar{\psi}(x_{3})\bar{\Phi}(x_{4})S]|0>
\end{equation}
 where
 \begin{equation}
 S=\sum \frac{(-i)^{n}}{n!}\int dy_{1}\ldots dy_{n}T\left\{ \mathcal{H}_{int}(y_{1})\ldots \mathcal{H}_{int}(y_{n})\right\}
 \end{equation}
 The primes indicate that the fields are the fully interacting fields in the Heisenberg representation. $G_{0}$ is the Green's function for non-interacting fields (i.e., $\Phi$'s and $\psi$'s are not interacting with each other but the electromagnetic field is not turned off) and $I$ is the irreducible interacting kernel which is defined by Eq. \ref{BSeq}. Using Eq. \ref{coulomb} in the Bethe-Salpeter equation, Eq. \ref{BSeq} we find for the Coulomb kernel
 
 \begin{equation}
 I_{c}(5,6;7,8)=-i\frac{e_{a}e_{b}}{4\pi}\gamma_{0}\delta(x_{5}-x_{7})\delta(x_{6}-x_{8})\delta(x^{0}_{5}-x^{0}_{6})\int d^{3}x'd^{3}y'\frac{\rho_{N}(x_{7}-x')\rho_{\pi}(x_{8}-y')}{|\mathbf{x}'-\mathbf{y}'|}
 \end{equation}
 and similarly this can be done for the other terms appearing in Eq.\ref{intH1}. Except for the appearance of the charge densities, $\rho_{N}$ and $\rho_{\pi}$ these are the same as appear in the paper by Owen \cite{foundations}.  So we shall not give them here.  The Bethe-Salpeter  equation appearing in Eq. \ref{BSeq} can be written in C. M. coordinates \footnote{for further details see reference \cite{owen}}.and for which the CM four vector $K=(K, \mathbf{0})$                                                      
 
 \begin{equation} \label{Green'sCM}
 \phi_{K}(x)=G_{K}(x,x')I_{K}(x',x'')\phi_{K}(x'')\ \ \ \ \ \textrm{where}\ \ \ \ \ \phi_{K}=<0|T\left\{\psi'(\eta_{b}x)\Phi'(-\eta_{a}x)\right\}|K>
\end{equation} 
 with $x=x_{1}-x_{2}$ and the only restriction on the $\eta$'s is that $\eta_{a}+\eta_{b}=1$.  
 Although $G_{K}(x,x')$ represents the free propagation of a nucleon and scalar meson in which each of the particles is fully 'dressed', the order $\alpha^{4}$ which is the accuracy of our calculation, we can treat each of these as free, non-interacting particles.  The contributions of the self-interactions does not contribute to order $\alpha^{4}$ and at most, can contribute to order $\alpha^{5}$.  Hence we can write $G_{K}(x,x')$ as \footnote{We have modified the definition of$ \hat{\Delta}(p)$ from what appears in \cite{foundations} by a factor $\beta_{s}$ so it now reads $\hat{\Delta}(p)=[2E(p)(p_{0}-\frac{M}{2m}-\beta_{s}m)]^{-1}\hat{\beta_{s}}$. This insures the correct statistics without the worrisome overall factor of $(2p_{0})^{-1}$ which was used in \cite{foundations}}
 
 \begin{eqnarray} \label{GreensF}
 G_{K}(x,x')&=&\int\frac{d^{4}p}{(2\pi)^{4}}\exp[-ip\cdot(x-x')]S_{F_{a}}(\eta_{a}K+p)\hat{\Delta}(\eta_{b}K-p) \nonumber \\
 &=&\int\frac{d^{4}p}{(2\pi)^{4}}\exp[-ip\cdot(x-x')]\frac{1}{2E_{b}(p)}\frac{1}{(\eta_{a}K+p_{0})\gamma_{0}-\mathbf{\gamma \cdot p}-m_{a}} \nonumber \\ & &\cdot\beta_{s}\frac{1}{(\eta_{b}K-p_{0})-\frac{M\mathbf{\gamma \cdot p}}{2m_{b}}-\beta_{s}m_{b}}
 \end{eqnarray}
 
 where
 \begin{eqnarray}
\hat{ \beta}_{s}=
\left( \begin{array}{cc} \frac{m_{b} +\frac{\mathbf{p}^{2}}{2m_{b}}}{E_{b}(p)} & \frac{\mathbf{p}^{2}}{2m_{b}E_{b}(p)} \\
-\frac{\mathbf{p}^{2}}{2m_{b}E_{b}(p)} & - \frac{m_{b} +\frac{\mathbf{p}^{2}}{2m_{b}}}{E_{b}(p)}
\end{array}\right) \approx \beta_{s}+\vartheta\left(\frac{\mathbf{p}^{4}}{m_{b}^{4}}\right)
 \end{eqnarray}
 
 By expanding this in partial fractions, we can write
 \begin{equation} \label{GreensF2}
 G_{K}(x,x')=\int\frac{d^{4}p}{(2\pi)^{4}}\exp[-ip\cdot (x-x')]\frac{1}{K-H_{a}-H_{b}}\Lambda_{K}(x,x')\hat{\beta}_{s}\gamma_{0}
\end{equation} 

with
\begin{equation}
\Lambda_{K}(x,x')=\int\frac{d^{4}p}{(2\pi)^{4}}\frac{e^{-ip\cdot(x-x')}}{2E(p)}\left\{\frac{1}{\eta_{a}K+p_{0}-\mathbf{\alpha \cdot p}-\beta m_{a}}+\frac{1}{\eta_{b}K-p_{0}-M\frac{\mathbf{p}^{2}}{2m_{b}}-\beta_{s}m_{b}}\right\}
\end{equation}

Using Eq. \ref{GreensF} in the Bethe-Salpeter equation in CM coordinates, Eq. \ref{Green'sCM},
we can write
\begin{eqnarray}
[K-\mathbf{\alpha \cdot p}-m_{a}\beta-M\frac{\mathbf{{p}^{2}}}{2m_{b}}-m_{b}\beta_{s}]\phi_{K}(x)&=&[\delta(x-x')+\Omega_{K}(x,x']\hat{\beta}_{s}\gamma_{0}i\frac{e_{a}e_{b}}{4\pi}\delta(x'-x'')\delta(x'_{0}) \nonumber \\
& &\cdot \int d^{3}u d^{3}v \frac{\rho_{N}(\mathbf{u})\rho_{\pi}(\mathbf{v})}{|\mathbf{x+u-v}|}\phi_{K}(x'') \nonumber \\
& &+I'_{K}(x,x') \phi_{K}(x')
\end{eqnarray}

where we have written $\Lambda_{K}(x,x')$ appearing in Eq.\ref{GreensF2}
\begin{equation}
\Lambda_{K}(x,x')=\delta(x-x')+\Omega_{K}(x,x')
\end{equation}

 This will enable us to obtain a Coulomb equation that is tractable and we have defined
 \begin{equation}
 I'_{K}(x,x')=I_{K}(x,x')-I_{c}(x,x')
 \end{equation}
 
In this work, we shall only be concerned with contributions of order $\alpha^{4}$ and since it has been shown that the term $\Omega_{K}(x,x')$ contributes are at most, order $\alpha^{5}$ so we can ignore this term as well.  Thus, if we can treat $I'_{K}(x,x')$ as a perturbation to the zero-order equation obtained by truncating $I'_{K}$ from the kernel.  Using our above treatment and the Foldy-Wouthuysen \cite{foldy} \cite{foundations} our zero-order equation is

\begin{equation} \label{Schrodinger}
\left[-\frac{\nabla^{2}}{2\mu}-\alpha\int d^{3}ud^{3}v \frac{\rho_{N}(\mathbf{u})\rho_{\pi}(\mathbf{v})}{|\mathbf{x+u+v}|}\right]\phi_{K_{c}}(x)=E\phi_{K_{c}}(\mathbf{x})
 \end{equation}
 
 where $E=K_{c}-m_{a}-m_{b}$ and $\mu$ is the \vspace{0.2in} reduced mass.

     The corrections to $E$ we designate by $\Delta E$ where $\Delta E=K-K_{c}$.   To order $\alpha^{4}$ (i.e., including all recoil contributions) we need to calculate both the relativistic  kinetic corrections (i.e., $E(p)-\frac{\mathbf{p}^{2}}{2m}$ and those arising from the one photon exchange between the proton and the bound meson.  From \cite{foundations} and from the treatment above, these are
     
 \begin{eqnarray} \label{delE}
\Delta E=&<&-\frac{\mathbf{p}^{4}}{8m_{a}^{3}}-\frac{\mathbf{p}^{4}}{8m_{b}^{3}}+\frac{\alpha}{4m_{a}^{2}}\int d^{3}ud^{3}v \rho_{N}(\mathbf{u})\rho_{\pi}(\mathbf{v})\left\{\frac{\mathbf{\sigma}_{a}\cdot (\mathbf{r+u-v})\times \mathbf{p}}{|\mathbf{r+u-v}|^{3}}+\frac{\alpha \pi}{2m_{a}^{2}}\delta(\mathbf{r+u-v})\right\}  \nonumber \\
&-&\frac{\alpha}{2m_{a}m_{b}}\int d^{3}u d^{3}v \rho_{N}(\mathbf{u}) \rho_{\pi}(\mathbf{v})\left\{ \frac{1}{|\mathbf{r+u-v}|}\mathbf{p}^{2}+\frac{\mathbf{r+u-v}}{|\mathbf{r+u-v}|^{3}}((\mathbf{r+u-v})\cdot \mathbf{p})\mathbf{p} \right. \nonumber \\
&-& \left.\frac{1}{|\mathbf{r+u-v}|^{3}}\mathbf{\sigma}_{a}\cdot (\mathbf{r+u-v})\times \mathbf{p}\right\}>
 \end{eqnarray}
     
   The first two terms in Eq. \ref{delE} are the relativistic kinetic terms, the next group of terms in the curly brackets are from the spin-orbit interaction and the terms in the last bracket are from the one-photon exchange.
   
   \section{Calculations}
   \subsection{Gaussian Charge Distribution}
   One could assume, as a reasonable approximation to take both $\rho_{N}$ and $\rho_{\pi}$ as Gaussian.  That is
   
  \begin{equation} \label{densities}
  \rho_{N}(\mathbf{u},\sigma_{N})=\frac{1}{\sqrt{(2\pi)^{3}}\sigma_{N}^{6}}\exp(-\frac{\mathbf{u}^{2}}{2\sigma_{N}^{2}}) \ \ \ \ \ \ \ \ \ \rho_{\pi}(\mathbf{v},\sigma_{\pi})=\frac{1}{\sqrt{(2\pi)^{3}}\sigma_{\pi}^{6}}\exp(-\frac{\mathbf{v}^{2}}{2\sigma_{\pi}^{2}})
 \end{equation}
 
 where $\sqrt{<r^{2}>_{N}}=\sqrt{3\sigma_{N}^{2}}$ and $\sqrt{<r^{2}>_{\pi}}=\sqrt{3\sigma_{\pi}^{2}}$.  However, as a first approximation we take the simpler case for which the meson is point-like, i.e., $\sigma_{\pi}\rightarrow 0$.  Then $\rho_{\pi}(\mathbf{v},\sigma_{\pi})=\delta(\mathbf{v})$.  The integral in Eq. \ref{Schrodinger} reduces to

 \[ \int d^{3}u \frac{\rho_{N}(\mathbf{u})}{|\mathbf{r+u}|}    \]
 
 and for $\rho_{N}$ in Eq. \ref{densities} we have a rather simple result
 
 \begin{equation}
 \int d^{3}u\frac{\rho_{N}(\mathbf{u},\sigma_{N})}{|\mathbf{r+u}|}=\frac{1}{r}\textrm{erf}\left[\frac{r}{\sqrt{2}\sigma_{N}}\right]
 \end{equation}
 
 Writing the Coulomb wave function appearing in Eq. \ref{Schrodinger} as $\phi_{K_{c}}(\mathbf{r})=R_{nl}(r)Y_{lm}(\theta,\phi)$ the equation for $R_{nl}(r)$ following from Eq. \ref{Schrodinger} is
 
 \begin{equation} \label{Schroedinger2}
 \frac{1}{r^{2}}\frac{d}{dr}r^{2}\frac{dR_{nl}(r)}{dr}-\frac{l(l+1)}{r^{2}}R_{nl}(r)+2\mu\left(E_{nl}+\alpha\frac{1}{r}\textrm{erf}\left[\frac{r}{\sqrt{2}\sigma_{N}}\right]\right)R_{nl}(r)=0
 \end{equation}
 
 For this approximation in which we take the meson radius to be zero and the nucleon rms to be $\sqrt{3\sigma^{2}_{N}}$ we need to solve Eq. \ref{Schroedinger2} and use the $\phi_{K_{c}}(r)$ that we obtain to evaluate the contribution of the terms Eq. \ref{delE} where we can take $\mathbf{v}=0$ since $\rho_{\pi}(\mathbf{v})=\delta(\mathbf{v})$. For this case, $\Delta E$ of Eq \ref{delE} becomes
 
 \begin{eqnarray} \label{delE2}
 \Delta E= &-&\left< \frac{\mathbf{p}^{4}}{8m_{a}^{3}}-\frac{\mathbf{p}^{4}}{8m_{b}^{3}}+\frac{\alpha}{4m_{a}^{2}r^{2}}\left(\frac{1}{r}\textrm{erf}(\frac{r}{\sqrt{2}\sigma_{N}})-\frac{1}{\sqrt{2}\sigma_{N}}e^{-\frac{r^{2}}{2\sigma_{N}^{2}}}\right)\mathbf{\sigma}_{a}\cdot \mathbf{r \times p} \right. \nonumber  \\ 
 &+&\frac{\alpha \pi}{2m_{a}^{2}} \frac{1}{\sqrt{(2\pi)^{3}\sigma_{N}^{6}}}e^{-\frac{r^{2}}{2\sigma_{N}}}-\frac{\alpha}{2m_{a}m_{b}r}\left(\textrm{erf}(\frac{r}{\sqrt{2}\sigma_{N}})\mathbf{p}^{2}+\frac{1}{r^{3}}\left[\sqrt{\frac{8}{\pi}}\sigma_{N}e^{-\frac{r^{2}}{2\sigma_{N}^{2}}} \right.\right. \nonumber \\
 &+&\left.\frac{1}{r}(r^{2}-2\sigma_{N}^{2})\textrm{erf}(\frac{r}{\sqrt{2}\sigma_{N}})\right]\mathbf{r\cdot (r\cdot p)p}-\frac{1}{r^{2}}\left[\frac{1}{r}\textrm{erf}(\frac{r}{\sqrt{2}\sigma_{N}})-\frac{1}{\sqrt{2}\sigma_{N}}e^{-\frac{r^{2}}{2\sigma_{N}^{2}}}\right] \nonumber \\
& &\left. \left. \cdot \mathbf{\sigma}_{a}\cdot \mathbf{r \times p}\right)\right>
\end{eqnarray}
 
 where the brackets mean that the values are to be calculated from the wave function obtained by solving Eq \ref{Schroedinger2}.
 
 The following tables include the $\alpha^{4}$ results in eV calculated from Eq.\ref{delE2}\footnote{The nuclear RMS used for the following tables was taken to be $1.2\sqrt{3/5}A^{1/3}$fm where A is the atomic number}. 
 
 \begin{tabular}{||c|r|r|r||} \hline
 contributions & kaonic hydrogen & kaonic helium$^{3}$  & kaonic carbon$^{13}$   \\ \hline \hline
 $\frac{(Z\alpha)^{2}\mu}{2}$ & 8616.7 & 44739.9  & 45876 \\ \hline
 finite size correction & -1.24 & -295.3 & -61220 \\  \hline \hline
 total $(Z\alpha)^{2}$ contribution & 8615.5  & 44444.6 &  393656 \\   \hline \hline
 \end{tabular}
 
 Energy Contributions  \vspace{0.25in}of order $(Z\alpha)^{2}$

 \begin{tabular}{||c|r|r|r||} \hline
  contributions & kaonic hydrogen & kaonic helium$^{3}$  & kaonic carbon$^{13}$   \\ \hline \hline
  $-\frac{<p^{4}>}{8m_{1}}$ & -0.0302& -6.0069 & -403.46  \\  \hline
   $-\frac{<p^{4}>}{8m_{2}}$ &-0.0044 & 0.0326 & -0.03     \\ \hline
   $<\frac{Z\alpha \pi}{2m_{1}^{2}\sqrt{(2\pi)^{3}}\sigma^{6}}\exp{(-\frac{r^{2}}{2\sigma^{2}})}>$& -0.0531 & -0.1699 & -0.50 \\  \hline
   $-<\frac{Z\alpha}{2m_{1}m_{2}r}\textrm{erf}(\frac{r}{\sqrt{2}\sigma})\mathbf{p}^{2}>$&14.4623 & 79.7487 & 661.73  \\  \hline
   $-<\frac{Z\alpha}{2m_{1}m_{2}r}\sqrt{\frac{8}{\pi}}\exp{(-\frac{r^{2}}{2\sigma^{2}})}\mathbf{r\cdot(r\cdot p)p}>$& 0.0011 & 0.1326 & 10.84  \\ \hline
   $-<\frac{Z\alpha}{2m_{1}m_{2}r}(\frac{1}{r^{2}}-\frac{2\sigma^{2}}{r^{4}})\textrm{erf}(\frac{r}{\sqrt{2}\sigma})\mathbf{r\cdot(r\cdot p)p}>$ & 0.0983 & 0.7602 & 0.28 \\  \hline  \hline
   total $(Z\alpha)^{4}$ contribution & 14.474 & 74.4321 & 268.31  \\ \hline  \hline
 \end{tabular}
 Energy Corrections order $(Z\alpha)^{4}$
 
 \section{Conclusion}
 We have used the Bethe-Salpeter equation to calculate to $(Z\alpha)^{4}$ electromagnetic contribution to the energy levels of kaonic hydrogen, kaonic helium ($K-^{3}He$) and kaonic carbon ($K-^{13}C$).  We have included finite size effects by assuming a gaussian charge distribution in the heavier particle.  To simplify the calculation we neglect the finite size of kaon and calculate the much larger effect of the nuclear charge distribution.  As can be seen in the second table, these contributions are not negligible.  All recoil contributions are obtained exactly by the use of the Bethe-Salpeter equation.  The knowledge of the electromagnetic contribution to these energy levels is essential to untangle the strong interaction from that of the electromagnetic contribution \cite{deloff}.  Furthermore the investigation of bound state systems such as these, because the interaction time is much longer than that in scattering experiments, allows for additional, precision tests of QCD.

\end{document}